\RequirePackage{fix-cm}
\documentclass[smallextended]{svjour3}       
\smartqed  
\usepackage{graphicx}
\usepackage{amssymb,amsmath}
\usepackage{bm}
\usepackage{color}
 \journalname{Journal of Low Temperature Physics}
\begin{document}

\title{Momentum distribution of liquid $^{\bf 4}$He across the normal-superfluid phase transition
}

\titlerunning{Momentum distribution of liquid $^{\bf 4}$He across
$T_\lambda$}        

\author{G. Ferr\'e         \and
        R. Rota \and J. Boronat 
}


\institute{G. Ferr\'e and J. Boronat\at
              Departament de F\'\i sica, Universitat Polit\`ecnica de Catalunya \\
              E-08034 Barcelona, Spain\\
           \and
           R. Rota \at
           Laboratoire Mat\'eriaux et Ph\'enom\`enes Quantiques, \\ 
              Universit\'e Paris Diderot, Sorbonne Paris Cit\'e, \\ 
CNRS-UMR7162, 75013 Paris, France
}

\date{Received: date / Accepted: date}

\maketitle

\begin{abstract}
We have carried out a study of the momentum distribution and of the spectrum of elementary excitations of liquid $^4$He
across the normal-superfluid transition temperature, using the path integral 
Monte Carlo method. Our results for the momentum distribution in the superfluid regime show that a kink 
is present in the range of momenta corresponding to the roton excitation. This effect disappears when crossing the transition temperature to the normal fluid, in a behavior currently unexplained by theory.

\end{abstract}

\section{Introduction}
\label{intro}

Continued theoretical and experimental work in the last decades have
led to an accurate knowledge of the superfluid transition in liquid $^4$He ~\cite{glyde}.
Probably, the most important feature is the suppression of the superfluid
density above the critical temperature $T_\lambda=2.17$ K. The
condensate fraction, which quantifies the macroscopic occupation of the
zero-momentum state, vanishes at the same point. The nature of the
excitations of the system is also modified in a significant way. By means of
neutron scattering one can access to the dynamic structure function $S(\bm{
k},\omega)$ which contains the maximum attainable information on the
excitations of the fluid~\cite{lovesey}. The most noticeable feature in the
dynamic response when $T_\lambda$ is crossed is the disappearance of the
roton as a quasi-particle mode and, as a consequence, of the conditions for superfluidity according to the Landau criterion.

The behavior of the momentum distribution of the $^4$He atoms $n({\bf k})$ is notably different in the two sides of the transition, the change being mainly in the limit of low momenta. First of all, in the superfluid regime, the presence of a condensate gives rise to a sharp delta contribution at ${\bf k}=0$. Furthermore, $n({\bf k})$ shows a singular $1/k$ behavior which can be explained in terms of a coupling between the condensate and the long-wavelength excitations (phonons) \cite{gavoret,sokol}. On the other hand, the width of $n({\bf k})$ increases slightly with $T$ by a merely thermal effect. However, some theoretical calculations in the limit of zero temperature point out that the shape of $n({\bf k})$ presents a change in the slope, or a kink, at $k \simeq 2$ \AA$^{-1}$ ~\cite{pandha,moroni,rota}, i.e. in the regime of momenta in which the dynamic structure factor displays a strong quasiparticle peak corresponding to the roton. This tiny effect was not studied in previous calculations at finite temperature \cite{ceperley86}. It is therefore plausible to think that the kink in $n({\bf k})$ can be related to the roton mode, in a similar way as the $1/k$ divergence at low momenta is ascribed to phonons.

The aim of this paper is to investigate the relation between the presence of the kink in $n({\bf
k})$ and that of the roton-mode in $S({\bf k},\omega)$. For this purpose, we perform path integral Monte Carlo
(PIMC) simulations of $^4$He across the normal-superfluid transition and we calculate both the momentum distribution and the
roton energy and strength. Our results do show that, as the liquid enters the normal phase, the kink is slowly smoothed out and the strength of the roton mode vanishes, supporting thus the hypothesis that the kink in the momentum distribution is a signal of the presence of the roton mode.        

In the next Section, we introduce some basics on the PIMC method and on the
stochastic optimization method used for the estimation of the dynamic
response from the imaginary-time correlation factor. In Sec. 3, we report
the results obtained on the momentum distribution and strength of the roton
mode in a range of temperatures across the normal-superfluid phase
transition. Finally, Sec. 4 comprises a brief summary of the paper and the
main conclusions resulting from our microscopic calculations.

\section{The path integral Monte Carlo method}
\label{method}

At the microscopic level, liquid $^{\bf 4}$He can be described as a quantum $N$-body system, whose Hamiltonian takes the form:
\begin{equation}
\label{eq:Hamiltonian}
\hat{H} = \hat{K} + \hat{V}= -\frac{\hbar^2}{2m} \sum_{i=1}^N \nabla^2 + \sum_{i<j} V(|{\bf r}_i - {\bf r}_j|) \ ,
\end{equation}
where $m$ is the mass of a $^4$He atom. The interaction between two particles in ${\bf r}_i$ and ${\bf r}_j$ is accurately described by the Aziz potential \cite{aziz}.

The properties of this system at a finite temperature $T$ are obtainable from the thermal density matrix $\hat{\rho} = {e^{-\beta \hat{H}}}/{Z}$,
where $\beta=1/(k_B T)$, $k_B$ is the Boltzmann constant, and $Z=\text{Tr} (e^{-\beta \hat{H}})$ is the partition
function. The knowledge of $\hat{\rho}$ allows for the calculation of the 
expected value of any operator $\hat{O}$,
\begin{equation}
\langle \hat{O} \rangle = \text{Tr} (\hat{\rho} \, \hat{O}) = \int d \bm{R} \ \rho(\bm{R},\bm{R};\beta) O({\bf R}) \ ,
\label{eq:expected}
\end{equation} 
where in the last term of the equation we have used the coordinate representation, $\bm{R}=\{ \bm{r}_1,\ldots,\bm{r}_N\}$ being the set of coordinates of the $N$ particles in the system. Deep in the quantum regime, i.e. at very low temperature, the estimation of the thermal density matrix is unfeasible, due to the non-commutativity of the kinetic and potential energy operators appearing in the Hamiltonian (Eq. \ref{eq:Hamiltonian}). Nevertheless, this problem can be overcome applying the convolution property of $\hat{\rho}$ and rewriting Eq. \ref{eq:expected} as
\begin{equation}
\langle \hat{O} \rangle = \int d \bm{R}_1 \ldots d \bm{R}_M \
\prod_{j=1}^{M} \rho(\bm{R}_j,\bm{R}_{j+1};\tau) O(\bm{R}_1 \ldots \bm{R}_M)\ , 
\label{eq:convolution}
\end{equation}
with $M$ being an integer and $\tau=\beta/M$. In this way, the thermal density
matrix at the desired temperature $T$ is obtained from a product of density matrices at a
higher temperature $MT$, where the effects of the non commutativity of $\hat{K}$ and $\hat{V}$ are reduced and it is easy to build approximations for the quantum density
matrix. 

For a Bose system, the function $\rho(\bm{R}_j,\bm{R}_{j+1};\tau)$ is positive definite and the product appearing in Eq. \ref{eq:convolution} can be considered as a probability distribution. Thus, the thermal averages $\langle \hat{O} \rangle$ can be efficiently estimated with a stochastic Monte Carlo technique, the so-called path integral Monte Carlo (PIMC) method \cite{ceperleyRMP}. Although increasing the number $M$ of convolution terms (usually called \textit{beads}) in Eq. \ref{eq:convolution} allows to reduce the systematic error due to the approximation of the thermal density matrix and eventually to recover "exactly" the expectation values, it is fundamental to use a good approximation for the thermal density matrix in order to reduce the numerical complexity of the algorithm and to avoid ergodicity issues in the sampling. To this end, we use a fourth-order time-step ($\tau$) approximation ~\cite{kostas}, based on a symplectic expansion of the propagator due to Chin~\cite{chin}. 

An additional difficulty in the PIMC simulations of liquid $^4$He at low temperatures arises from the indistinguishability of the particles. This problem can be solved sampling numerically the bosonic permutations among the atoms: we have used the worm algorithm which is an efficient method aimed at this goal \cite{worm}.

\begin{figure}
\begin{center}
\includegraphics[width=0.70\linewidth,angle=0]{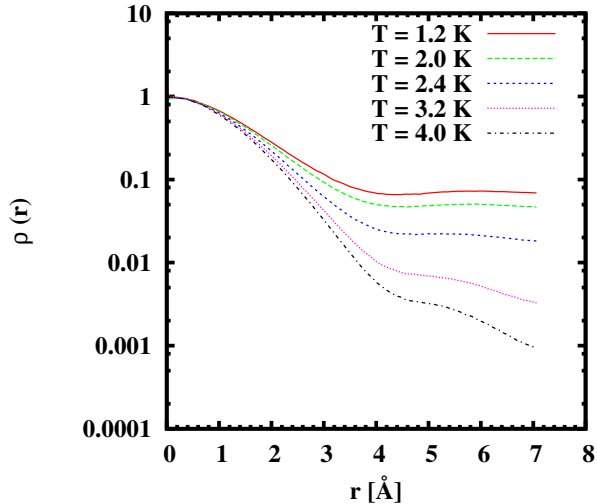}
\caption{(Color online) One-body density matrix of liquid $^4$He at 
different temperatures at saturated vapor pressure (SVP).
}
\label{Fig:1}
\end{center}
\end{figure}

The PIMC method is able to give an accurate description of the static properties of Bose systems, but it does not allow for the calculations of real-time correlation functions, which are fundamental to study dynamic properties, such as the spectrum of the elementary excitations. A quantity that is easily accessible in PIMC simulation is the intermediate scattering function $F(\bm{k},\tau) = 1/N \, \langle \hat{\rho}_{\bm{k}}(\tau)
\, \hat{\rho}_{\bm{k}}^\dagger(0) \rangle$, with $\hat{\rho}_{\bm{k}}(\tau)=\sum_{i=1}^N e^{i \bm{k} \cdot \bm{r}_i}$
being the density fluctuation operator. This function $F(\bm{k},\tau)$ can be considered as a correlation function in imaginary time and it can be related to the dynamic structure factor $S(\bm{k},\omega)$ by a Laplace Transform.
Even if the inversion of the Laplace transform is an ill-posed problem, many numerical algorithms have been developed to recover reasonable estimates of $S(\bm{k},\omega)$ from the PIMC data for $F(\bm{k},\tau)$ \cite{jarrellME,sandwik,mishckenko,prokofev13,vitali}. In this work, to obtain the dynamic structure factor, we use a stochastic optimization algorithm based on simulated annealing,  as described in Ref. \cite{ferre}.

\section{Results}
\label{results}

We have performed PIMC calculations of liquid $^{\bf 4}$He following the
SVP and $p = 10 \ atm$ densities, from $T = 0.8$ to $4 \mathrm{K}$, 
with $N = 64$ particles 
in the simulation box under periodic boundary conditions. We have 
checked that the use of larger number of particles does not
modify the main results discussed in this Section.

Following Eq. \ref{eq:convolution}, we implemented the calculation of the 
one-body density matrix, defined as
\begin{equation}
\rho_1(\bm{r}_1,\bm{r'}_1) = \frac{V}{Z} \, \int d \bm{r}_2 \ldots d \bm{r}_M \rho(\bm{R},\bm{R'};\beta) \ ,
\label{eq:onebody}
\end{equation}
where $\rho(\bm{R},\bm{R'};\beta)$ is the thermal density matrix 
computed for two configurations $\bm{R}=\{ \bm{r}_1, \bm{r}_2 \ldots,\bm{r}_N\}$ 
 and
$\bm{R'}=\{ \bm{r'}_1, \bm{r}_2 \ldots,\bm{r}_N\}$ which differs for 
the position of only one particle. 
We reported the influence of the temperature on this function in Fig. \ref{Fig:1}, sampled 
from non-diagonal configurations along the simulation.

As it is well known, there are significant differences between results for
$\rho_1(r)$ obtained below and above the critical temperature $T_\lambda$.
In the superfluid regime, $T<T_\lambda$, the one-body density matrix shows
a plateau at large distances corresponding to the presence of a finite
occupation of the zero-momentum state. Instead, in the normal phase,
$T>T_\lambda$, the one-body density matrix decays exponentially to zero pointing to the
absence of off-diagonal long range order in the system. 

\begin{figure}
\begin{center}
\includegraphics[width=1.0\linewidth,angle=0]{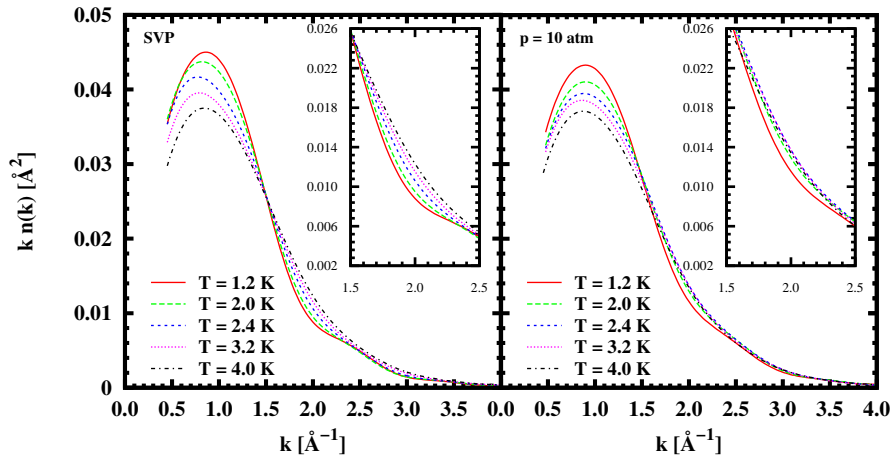}
\caption{(Color online) The momentum distribution plotted as 
$k n(k)$ at two different pressures and temperatures. The insets show the same
results but focused around $k \simeq 2$ \AA$^{-1}$. 
\textit{Left:} Results at saturated vapor pressure (SVP). 
\textit{Right:} Results at a higher pressure $p = 10$ atm.}

\label{Fig:2}
\end{center}
\end{figure}

\begin{figure}
\begin{center}
\includegraphics[width=1.0\linewidth,angle=0]{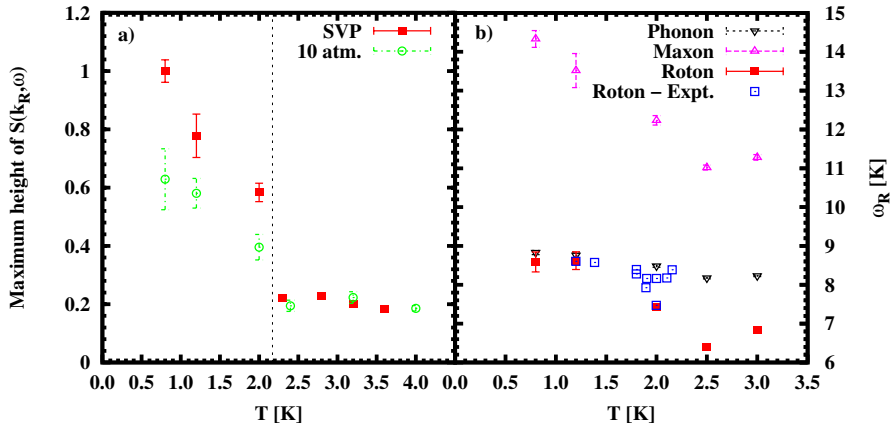}
\caption{(Color online) 
\textit{Left:} Maximum height of the dynamic structure factor 
$S(\bm{k},\omega)$ of the roton ($k \sim 1.91$ \AA$^{-1}$) across the 
normal-superfluid phase transition. All data are normalized 
by the maximum height at $T=0.8$ K and SVP.
\textit{Right:} Temperature dependence of the phonon ($k \sim 0.43$ \AA$^{-1}$), maxon ($k \sim 1.24$ \AA$^{-1}$) and
roton ($k \sim 1.91$ \AA$^{-1}$) energy. Experimental results for the roton energy from Ref. \cite{mezei}. In both figures $k_R$ and $\omega_R$ are the momentum and the energy of the roton, respectively.
}
\label{Fig:3}
\end{center}
\end{figure}

The momentum distribution $n(\bm{k})$ can be obtained from the Fourier transform 
of the one-body density matrix $\rho_1(\bm{r})$ (Eq. \ref{eq:onebody}) as
\begin{equation}
n(\bm{k}) = n_0 \delta(\bm{k}) + \rho \, \int d^3 r \, e^{i \bm{k} \cdot \bm{r}} \left( \rho_1(r) - n_0 \right) \ ,
\label{eq:momentum}
\end{equation}
where $\rho$ stands for the density of our system and 
$n_0 = \lim_{r \rightarrow \infty} \rho_1(r)$ is the condensate fraction. Results of $n(k)$ for a range of temperatures across $T_\lambda$ are
reported in Fig. \ref{Fig:2}, plotted as $k n(k)$. Our data start at a $k$ value compatible with the finite size of the system,
$k_{\text{min}}=2 \pi/L$, with $L$ the length of any side of the cubic
simulation box. Therefore, we are not able to 
show the $k \to 0$ behavior of the momentum distribution.

When $T$ increases we see a progressive broadening of the distribution due to a
\textit{classical} thermal effect. However, in this evolution with $T$ we
can observe a non-trivial effect that appears at intermediate $k$ values,
$1.5 < k < 2.5$ \AA$^{-1}$. As we show in Fig. \ref{Fig:2}, and in
particular in the inset, there is a kink of $n(k)$ within this $k$ range
for temperatures smaller than $T_\lambda$, i.e., in the superfluid regime.
As the temperature 
increases, and goes near the transition point, the kink becomes smoother, 
and it completely disappears for $T > T_\lambda$.  
We also notice that the kink is a bit more pronounced at SVP, when the 
intensity of the roton peak in the dynamic structure factor is larger (see Fig \ref{Fig:3}a).
The location of this kink around $k\simeq 2$ \AA$^{-1}$
leads us to think that the kink can be related to 
the characteristic momentum of the roton excitation. It is known that the
roton quasi-particle excitation is associated to the superfluidity of the
system through the Landau criterium. In the normal phase, the roton
disappears as a quasi-particle peak in the dynamic response $S({\bm k},\omega)$.
Therefore, the connection between this kink in $n(k)$ and the roton
excitation seems rather plausible.

The evolution of the roton with temperature has been studied by means
of the inverse Laplace transform of the intermediate scattering
function~\cite{ferre},
as commented in the previous Section.
In Fig. \ref{Fig:3} (left panel), we show how the maximum height 
of the dynamic structure factor slowly decreases in the superfluid 
phase as we increase the temperature, until it experiences an abrupt drop 
once we enter the normal phase and then on it remains constant. 
This is an expected result since the 
quasi-particle peak of the roton excitation disappears once 
we cross $T_\lambda$. Our data, reported in the figure, also show that the
strength of the roton peak is slightly reduced when the pressure increases.

We can also look for the energy of the roton excitation, as well as
the phonon and maxon excitations, and see how they evolve with  
temperature (see Fig. \ref{Fig:3}, right panel). In the superfluid phase, 
for the roton, the energy decreases as we increase the temperature.
At higher temperatures, in the normal phase,  it seems 
that the energy raises again, but one can not really speak 
about roton mode
anymore due to the wide broadening of its peak. 
As a matter of comparison, we report in Fig. \ref{Fig:3} data obtained for
the maxon and phonon energies.
For the maxon excitation, the behavior
is similar to the one of the roton, but in this case the strength of the 
peak when crossing
$T_\lambda$ is not so drastically reduced~\cite{ferre}.  
In the case of the phonon, 
the influence of the temperature is much smaller than in the previous cases.

\section{Conclusions}
We have carried out a microscopic study of the momentum distribution of
liquid $^4$He at finite temperature using the PIMC method. Our aim has been
to determine the possible origin of the kink that $n(k)$ shows at $k$
values around the roton momentum. This is not the first observation of this
kink in theoretical calculations since it was already obtained more that
twenty years ago~\cite{pandha}. The location of the kink around the roton momentum led
the idea of its relation with the roton but without further analysis. Now,
we have shown that this scenario is more plausible because the kink
vanishes when $T_\lambda$ is crossed, mimicking the behavior of the roton
quasi-particle peak.

\begin{acknowledgements}
We acknowledge partial financial support from the MICINN (Spain) Grant No.~FIS2014-56257-C2-1-P.
\end{acknowledgements}



\end{document}